\documentclass[12pt]{iopart}

\usepackage{iopams}
\usepackage{graphicx}
\newcommand{\nn}{\nonumber}

\newcommand{\hata}{\hat{a}}
\newcommand{\hatad}{\hat{a}^\dag}

\newcommand{\hatAd}{\hat{A}^\dag}

\newcommand{\la}{\langle}
\newcommand{\ra}{\rangle}

\newcommand{\Uh}{\hat{U}}

\begin{document}
%
\title[Quantum propagator for a general time-dependent quadratic Hamiltonian]{Quantum propagator for a general time-dependent quadratic Hamiltonian: Application to interacting oscillators in external fields}

\author{Shohreh Janjan}
\address{Department of Physics, University of Kurdistan, Sanandaj, P.O.Box 66177-15175}
\ead{sh.janjan@uok.ac.ir} 

\author{Fardin Kheirandish}
\address{Department of Physics, University of Kurdistan, Sanandaj, P.O.Box 66177-15175}
\ead{f.kheirandish@uok.ac.ir}


\date{\today}
\begin{abstract}
\noindent In this paper, we find the quantum propagator for a general time-dependent quadratic Hamiltonian. The method is based on the properties of the propagator and the fact that the quantum propagator fulfills two independent partial differential equations originating from Heisenberg equations for positions and momenta. As an application of the method, we find the quantum propagator for a linear chain of interacting oscillators for both periodic and Dirichlet boundary conditions. The state and excitation propagation along the harmonic chain in the absence and presence of an external classical source is studied and discussed. The location of the first maxima of the probability amplitude $P(n,\tau)$ is a straight line in the $(n,\tau)$-plane, indicating a constant speed of excitation propagation along the chain.
\end{abstract}

\maketitle 

\section{Introduction}
\noindent An important area of quantum physics is the study of the properties and applications of interacting harmonic oscillators. This interest in such systems is due to their wide applications in quantum and nonlinear physics \cite{1fano1957description,2han1990linear,3iachello1991model,4prauzner2004two,5paz2008dynamics,6galve2010bringing}, chemistry \cite{7ikeda1999incoherent,8delor2017directing}, and biophysics \cite{9fuller2014vibronic,10halpin2014two}. The coupled harmonic oscillators are also important in quantum optics, a linear beam splitter in quantum optics can be modelled by two coupled harmonic oscillators \cite{11makarov2020quantum}. The bound harmonic oscillators are a good model of real physical objects such as thermal vibrations of coupled atoms, photons in cavities, ions in ion-traps, and many more. The study of coupled harmonic oscillators is also one of the main theoretical methods to investigate quantum decoherence \cite{12zurek1991quantum,13zurek2003decoherence,14schlosshauer2007decoherence} and entanglement \cite{15plenio2004dynamics,16makarov2018coupled,17audenaert2002entanglement,18makarov2020quantum}. In non-relativistic quantum mechanics, propagators can be calculated in three different approaches \cite{19beauregard1966propagators}: i) Using the explicit form of the energy eigenstates satisfying predefined boundary conditions. ii) Using path integral techniques \cite{20feynman2010quantum,21schulman2012techniques}. iii) Solving the Heisenberg equations of motion for the problem in question \cite{22kheirandish2018exact,23kheirandish2018novel}. The Van Vleck Formula, Maslov Theory,
and the semiclassical approach has been investigated in \cite{littlejohn1992van,littlejohn1986semiclassical}, and for a metaplectic operator approach see \cite{lee2023metaplectic,de2014metaplectic}. Here we follow the third approach to find a closed form propagator for a general time-dependent quadratic Hamiltonian influenced by an external classical source. The method is based on the properties of the propagator and the fact that the quantum propagator
fulfills two independent partial differential equations originating from Heisenberg equations for the positions and momenta operators. As an application of the method, we find the quantum propagator for a linear chain of interacting oscillators for both periodic and non-periodic boundary conditions. In the following, we study and discuss the state and excitation propagation along the harmonic chain in the absence and presence of an external classical source. We show that, the location of the first maxima of the probability amplitude $P(n,\tau)$ is a straight line in the $(n,\tau)$-plane, indicating a constant speed of excitation propagation along the chain.
\section{Hamiltonian}
\noindent In this section, we consider a general time-dependent integrable quadratic Hamiltonian defined by the Hamiltonian \cite{24robert2021coherent}
\begin{eqnarray}\label{1}
\hat{H}(t,\hat{q},\hat{p}) &=& \frac{1}{2}\hat{q}\cdot Z_t \cdot\hat{q}+\frac{1}{2}\hat{q}\cdot L^{T}_t\cdot\hat{p}+\frac{1}{2}\hat{p}\cdot L_t\cdot\hat{q} +\frac{1}{2}\hat{p}\cdot K_t\cdot\hat{p}\nonumber\\
&&-\mu(t)\cdot\hat{q}-\nu(t)\cdot\hat{p},\nonumber\\
                           &=& \frac{1}{2}
                           \left(
                             \begin{array}{cc}
                              \hat{q} & \hat{p} \\
                             \end{array}
                           \right)
                           w(t)
                           \left(
                             \begin{array}{c}
                              \hat{q} \\
                              \hat{q} \\
                             \end{array}
                           \right)
                           -
                           \left(\begin{array}{cc}
	                       \mu(t) & \nu(t)
                           \end{array}
                           \right)
                            \left(
                             \begin{array}{c}
                              \hat{q} \\
                              \hat{q} \\
                             \end{array}
                           \right),
\end{eqnarray}
where the super index $T$ denotes the transpose operation on a matrix and the column matrices $q=[q^1, q^2, \cdots, q^n]^T$ and $p=[p^1, p^2, \cdots, p^n]^T$ are $n$-dimensional position and momentum operators. The matrices $Z_t$ and $K_t$ and $L_t$ are $n\times n$ matrices with $Z_t$ and $K_t$ being symmetric. The $n$-dimensional classical time-dependent vectors $\mu(t)=[\mu^1(t), \mu^2 (t),\cdots,\mu^n (t)]^T$ and $\nu(t)=[\nu^1(t), \nu^2 (t),\cdots,\nu^n (t)]^T$ are external sources coupled to positions and momenta, respectively. The $2n\times 2n$ matrix $w(t)$ is defined by
\begin{equation}\label{2}
 w(t)=\left(
        \begin{array}{cc}
          Z_t & L^T_t \\
          L_t & K_t \\
        \end{array}
        \right).
\end{equation}

From Heisenberg equation $i\hbar\,\dot{\hat{q}}(t)=[\hat{q}(t),\hat{H}(t)]$, we find
\begin{eqnarray}\label{3}
\dot{\hat{q}}(t) =L_t\cdot\hat{q}(t)+K_t\cdot\hat{p}(t)-\nu(t),\label{3}
\end{eqnarray}
and similarly, from $i\hbar\,\dot{\hat{p}}(t)=[\hat{p}(t),\hat{H}(t)]$, we obtain
\begin{eqnarray}\label{4}
\dot{\hat{p}}(t) = -Z_t\cdot\hat{q}(t)-L^{T}_t\cdot\hat{p}(t)+\mu(t).\label{4}
\end{eqnarray}
The equations (\ref{3},\ref{4}) can be written in matrix form as
\begin{equation}\label{5}
\left(
\begin{array}{cc}
	\dot{\hat{q}}(t) \\
	\dot{\hat{p}}(t) \\
\end{array}
\right)
=
\left(
\begin{array}{cc}
	 L_t & K_t \\
	-Z_t & -L^{T}_t \\
\end{array}
\right)
\left(
\begin{array}{c}
	\hat{q}(t) \\
	\hat{p}(t) \\
\end{array}
\right)
+
\left(
\begin{array}{c}
	-\nu(t) \\
	\mu(t) \\
\end{array}
\right),
\end{equation}

The coefficient matrix in Eq. (\ref{5}) can be written as
\begin{equation}\label{5-2}
\left(
\begin{array}{cc}
	 L_t & K_t \\
	-Z_t & -L^{T}_t \\
\end{array}
\right)=
\left(
\begin{array}{cc}
	 0 & \mathbf{1}_n \\
	-\mathbf{1}_n & 0 \\
\end{array}
\right)
\left(\begin{array}{cc}
          Z_t & L^T_t \\
          L_t & K_t \\
        \end{array}
        \right)=s\cdot w(t),
\end{equation}
where the symplectic matrix $s$ is defined by
\begin{equation}\label{6}
 s=
 \left(\begin{array}{cc}
	 0 & \mathbf{1}_n \\
	-\mathbf{1}_n & 0 \\
\end{array}
\right),
\end{equation}
and $\mathbf{1}_n$ is the $n\times n$ identity matrix. Now Eq. (\ref{5}) can be written as
\begin{equation}\label{5-3}
\left(
\begin{array}{cc}
	\dot{\hat{q}}(t) \\
	\dot{\hat{p}}(t) \\
\end{array}
\right)
=
s\cdot w(t)
\left(
\begin{array}{c}
	\hat{q}(t) \\
	\hat{p}(t) \\
\end{array}
\right)
+
\left(
\begin{array}{c}
	-\nu(t) \\
	\mu(t) \\
\end{array}
\right).
\end{equation}
Let us define the $2n\times 2n$ matrix $\vartheta(t)$ as
\begin{equation}\label{7-0}
 \vartheta(t)
 =
 \left(
 \begin{array}{cc}
	 A_t & B_t \\
	C_t & D_t \\
\end{array}
\right),
\end{equation}
where $A_t $, $B_t $, $C_t $, and $D_t $ are $n\times n$ unknown matrices. We assume that $\vartheta(t)$ fulfills the homogeneous part of Eq. (\ref{5-3})
\begin{equation}\label{7-5}
\Big(\mathbf{1}\partial_t-s\cdot w(t)\Big)\cdot\vartheta(t)=0,
\end{equation}
with initial condition $\vartheta(0)=\mathbf{1}_n$. By solving $\vartheta(t)$, the unknown matrices $A_t$, $B_t$, $C_t$, and $D_t$ will be determined. To solve Eq. (\ref{5-3}), we use the Green function technique. The matrix Green function $G(t,t')$ of Eq. (\ref{5-3}) fulfills the following equation
\begin{equation}\label{7-5}
\Big(\mathbf{1}_n\partial_t-s\cdot w(t)\Big)\cdot G(t-t')=\mathbf{1}_n\,\delta(t-t').
\end{equation}
By inserting the ansatz $G(t,t')=\theta(t-t')\vartheta(t)\cdot\vartheta^{-1}(t')$, into Eq. (\ref{7-5}), one can easily show that $G(t,t')$ is a Green function for Eq. (\ref{5-3}). Therefore, the general solution of Eq. (\ref{5-3}) can be written as
\begin{eqnarray}\label{7-2}
\left(
\begin{array}{c}
	\hat{q}(t) \\
	\hat{p}(t) \\
\end{array}
\right)
&=&
\vartheta(t)\cdot
\left(
\begin{array}{c}
	\hat{q}(0) \\
	\hat{p}(0) \\
\end{array}
\right)
+
\int_0^t \, dt'\,\vartheta(t)\cdot\vartheta^{-1}(t')
\cdot
\left(
\begin{array}{c}
	-\nu(t') \\
	\mu(t') \\
\end{array}
\right),\nn\\
&=&
\left(\begin{array}{cc}
	 A_t & B_t \\
	C_t & D_t \\
\end{array}
\right)\cdot
\left(
\begin{array}{c}
	\hat{q}(0) \\
	\hat{p}(0) \\
\end{array}
\right)
+
\left(
\begin{array}{c}
	\eta(t) \\
	\xi(t) \\
\end{array}
\right).
\end{eqnarray}
From now on, we assume that the homogeneous solution $\vartheta(t)$ is a known matrix, meaning that the $A_t $, $B_t $, $C_t $, and $D_t $ are all known time-dependent matrices. From Eq. (\ref{7-2}) we have
\begin{eqnarray}\label{qs}
&& \hat{q}(t)=A_t\cdot\hat{q}(0)+B_t\cdot\hat{p}(0)+\eta(t),\\
&& \hat{p}(t)=C_t\cdot\hat{q}(0)+D_t\cdot\hat{p}(0)+\xi(t).\label{ps}
\end{eqnarray}
The positions $\hat{q}(t)$ and momenta $\hat{p}(t)$ operators are Hermitian, so, the functions $\eta(t)$ and $\xi(t)$, and the matrices $A_t $, $B_t $, $C_t $, and $D_t $ are all real in any time. Also, from the canonical commutation relations $[\hat{q}(t),\hat{p}(t)]=\mathbf{1}_n\,i\hbar,$ one easily finds the relation $A_t\,D_t^T-B_t\,C_t^T=\mathbf{1}_n$.
\section{Quantum Propagator}
\noindent In this section for notational simplicity, we denote a classical $n$-component vector $(q_1,q_2,\cdots,q_n)$ by the bold face $\mathbf{q}$. To find the explicit form of the quantum propagator corresponding to the Hamiltonian Eq. (\ref{1}), we show that the quantum propagator $K(\mathbf{q},t|\mathbf{q}',0)$ fulfills two independent partial differential equations originating from Eqs. (\ref{qs},\ref{ps}). If we denote the time-evolution operator corresponding to Hamiltonian Eq. (\ref{1}) by $\hat{U}(t)$, then from Heisenberg picture we find
\begin{eqnarray}\label{qmain}
\hat{q}(t)=&& \hat{U}^{\dag}(t)\hat{q}(0)\hat{U}(t)=A_t\cdot\hat{q}(0)+B_t\cdot\hat{p}(0)+\eta(t),\\
\hat{p}(t)=&& \hat{U}^{\dag}(t)\hat{p}(0)\hat{U}(t)=C_t\cdot\hat{q}(0)+D_t\cdot\hat{p}(0)+\xi(t),\label{pmain}
\end{eqnarray}
or equivalently
\begin{eqnarray}\label{qmain2}
&&  \hat{U}(t)\,[A_t\cdot\hat{q}(0)+B_t\cdot\hat{p}(0)+\eta(t)] = \hat{q}(0)\hat{U}(t),\\
&&  \hat{U}(t)\,[C_t\cdot\hat{q}(0)+D_t\cdot\hat{p}(0)+\xi(t)] = \hat{p}(0)\hat{U}(t). \label{pmain2}
\end{eqnarray}
By applying the bra $\langle \mathbf{q}|$ on the left and the ket $|\mathbf{q}'\rangle$ on the right to both sides of Eqs. (\ref{qmain2},\ref{pmain2}), we find
\begin{eqnarray}\label{qmain3}
&& \langle \mathbf{q}| \hat{U}(t)\,[A_t\cdot\hat{q}(0)+B_t\cdot\hat{p}(0)+\eta(t)]|\mathbf{q}'\rangle = \langle \mathbf{q}|\hat{q}(0)\hat{U}(t)|\mathbf{q}'\rangle,\\
&&  \langle \mathbf{q}|\hat{U}(t)\,[C_t\cdot\hat{q}(0)+D_t\cdot\hat{p}(0)+\xi(t)]|\mathbf{q}'\rangle = \langle \mathbf{q}|\hat{p}(0)\hat{U}(t)|\mathbf{q}'\rangle. \label{pmain3}
\end{eqnarray}
In the position representation, Eqs. (\ref{qmain},\ref{pmain}) lead to the following partial differential equations
\begin{eqnarray}\label{qmain3}
&& \Big(B_t\cdot\nabla_{\mathbf{q}'}\Big)\ln K(\mathbf{q},t|\mathbf{q}',0) = -\frac{i}{\hbar}\Big(\mathbf{q}-\eta(t)-A_t\cdot \mathbf{q}'\Big),\\
&&  \Big(\nabla_\mathbf{q}+D_t\cdot\nabla_{\mathbf{q}'}\Big)\ln K(\mathbf{q},t|\mathbf{q}',0) = \frac{i}{\hbar}\Big(C_t\cdot \mathbf{q}'+\xi(t)\Big),\label{pmain3}
\end{eqnarray}
where $K(\mathbf{q},t|\mathbf{q}',0)=\la \mathbf{q}|\hat{U} (t)|\mathbf{q}'\ra$ is the kernel. From the right hand side of Eqs. (\ref{qmain3},\ref{pmain3}) we deduce that $\ln K(\mathbf{q},t|\mathbf{q}',0)$ is a quadratic function in terms of $\mathbf{q}'$ with coefficients that may depend on $\mathbf{q}$ or $t$. Therefore,
\begin{equation}\label{lnK}
  \ln K(\mathbf{q},t|\mathbf{q}',0)=f(\mathbf{q},t)+V(\mathbf{q},t)\cdot\mathbf{q}'+(1/2)\mathbf{q}'\cdot M(\mathbf{q},t)\cdot \mathbf{q}',
\end{equation}
Now by inserting Eq. (\ref{lnK}) into Eq. (\ref{qmain3}), we find $M(\mathbf{q},t)=(i/\hbar)\,B^{-1}_t\cdot A_t$ and $V(\mathbf{q},t)=(-i/\hbar)\,B^{-1}_t\cdot(\mathbf{q}-\eta(t))$. Also, by inserting Eq. (\ref{lnK}) into Eq. (\ref{pmain3}), we find
\begin{equation}\label{gradf}
  \nabla_\mathbf{q} f(\mathbf{q},t)=\frac{i}{\hbar}(\xi+D_t\cdot B_t^{-1}\cdot \mathbf{q}-D_t\cdot B_t^{-1}\cdot\eta).
\end{equation}
Therefore, $f(\mathbf{q},t)$ is quadratic in $\mathbf{q}$, and can be written as
\begin{equation}\label{fform}
  f(\mathbf{q},t)=f_0 (t)+W_t\cdot \mathbf{q}+\frac{1}{2}\mathbf{q}\cdot N_t\cdot \mathbf{q}.
\end{equation}
By inserting Eq. (\ref{fform}) into Eq. (\ref{gradf}), we find $W_t=\frac{i}{\hbar}(\xi-D_t\cdot B_t^{-1}\cdot \eta)$ and $N_t=\frac{i}{\hbar}\,D_t\cdot B_t^{-1}$. Now putting everything together we can write the propagator as
\begin{eqnarray}\label{10}
K(\mathbf{q},t|\mathbf{q}',0)=&& \Gamma(t)\,e^{\frac{i}{\hbar}\left[\frac{1}{2}\mathbf{q}\cdot D_t\cdot B^{-1}_t\cdot \mathbf{q}+\frac{1}{2}\mathbf{q}'\cdot B^{-1}_t\cdot A_t\cdot \mathbf{q}'-\mathbf{q}'\cdot B^{-1}_t\cdot \mathbf{q}\right]}\nn\\
&\cdot &\,e^{\left[\mathbf{q}\cdot(\xi(t)-D_t\cdot B^{-1}_t\cdot\eta(t))+\mathbf{q}'\cdot\eta(t)\right]}.
\end{eqnarray}
To find $\Gamma(t)$ we insert the propagator into the identity
\begin{eqnarray}\label{id1}
  \delta^n(\mathbf{q}-\mathbf{q}')&=& \langle \mathbf{q}|\hat{U}(t)\,\hat{U}^\dag (t)|\mathbf{q}'\rangle,\nn\\
              &=& \int d\mathbf{q}''\langle \mathbf{q}|\hat{U}(t)|\mathbf{q}''\rangle\,\langle \mathbf{q}''|\hat{U}^\dag(t)|\mathbf{q}'\rangle,\nn\\
              &=& \int d\mathbf{q}''\,K(q,t|\mathbf{q}'',0)\,\bar{K}(\mathbf{q}',t|\mathbf{q}'',0),
\end{eqnarray}
leading to
\begin{eqnarray}
   \delta^n(\mathbf{q}-\mathbf{q}') &=& |\Gamma(t)|^2\,(2\pi)^n\,\delta^n\left(\frac{1}{\hbar}\,B_t^{-1}\cdot (q-q')\right),\nn \\
                                  &=& |\Gamma(t)|^2\,(2\pi\hbar)^n\,(|\det B_t|)\, \delta^n(\mathbf{q}-\mathbf{q}'),
\end{eqnarray}
where we used the identity $\delta^n\big(B_t^{-1}\cdot (\mathbf{q}-\mathbf{q}')\big)=|\det(B_t)|\,\delta^n (\mathbf{q}-\mathbf{q}')$. Therefore,
\begin{equation}\label{At}
   \Gamma(t)=\left(\frac{1}{\sqrt{(2\pi\hbar)^n\,|\det B_t|}}\right)\,e^{i\theta(t)},
\end{equation}
where the phase $\theta(t)$ is a real function to be determined from the fact that the quantum propagator fulfills the Schr\"{o}dinger equation $i\hbar\,\partial_t K(\mathbf{q},t|\mathbf{q}',0)=\hat{H}K(\mathbf{q},t|\mathbf{q}',0)$ for arbitrary $\mathbf{q}'$. Therefore, we set $\mathbf{q}'=0$, and also after the partial differentiations with respect to $\mathbf{q}$, we will set $\mathbf{q}=0$, the final result is
\begin{equation}\label{f0}
  \frac{d\theta(t)}{dt}=-\frac{1}{2\hbar}\,\zeta\cdot K_t\cdot \zeta,
\end{equation}
where we made use of the identities
\begin{eqnarray}
B_t^{-1}\cdot \frac{dB_t}{dt}=L_t+K_t\cdot D_t\cdot B_t^{-1},\\
\frac{d }{dt}\det (B_t)=\det(B_t)\,\mbox{Tr} \left(B_t^{-1}\cdot \frac{dB_t}{dt}\right),
\end{eqnarray}
and for notational simplicity, we have defined $\zeta(t)=\xi_t-D_t\cdot B_t^{-1}\cdot \eta_t$. Therefore, $\theta(t)=\theta(0)-\frac{1}{2\hbar}\,\int_0^t dt'\,[\zeta\cdot K_t\cdot \zeta]_{t'}$. The constant phase $\theta(0)$ can be determine from  the property $\lim\limits_{t\rightarrow 0}\,K(\mathbf{q},t|\mathbf{q}',0)=\delta^n (\mathbf{q}-\mathbf{q}')$. For this purpose, using Eqs. (\ref{qmain},\ref{pmain}), we find $\lim\limits_{t\rightarrow 0} A_t=\lim\limits_{t\rightarrow 0} D_t=\mathbf{1}_n$ and $\lim\limits_{t\rightarrow 0} B_t=\lim\limits_{t\rightarrow 0} C_t=0$, and also using the following representation of the Dirac delta function \cite{li2013integral}
\begin{equation}\label{delta}
 \lim_{t\rightarrow 0} \left(\frac{\alpha}{\pi\, t }\right)^{\frac{1}{2}}\, e^{-\frac{\alpha (x-x')^2}{t}}=\delta(x-x'),
\end{equation}
we easily find $e^{i\theta(0)}=i^{-n/2}$. Finally, the exact form of the quantum propagator corresponding to the Hamiltonian Eq. (\ref{1}) is
\begin{eqnarray}\label{finalK}
K(\mathbf{q},t|\mathbf{q}',0) &=& \frac{1}{\sqrt{(2\pi i\hbar)^n\,|\det B_t}|}\,e^{-\frac{i}{2\hbar}\int_0^t dt'\,[\zeta\cdot K_t\cdot \zeta]_{t'}}\nonumber\\
&\cdot & e^{\frac{i}{\hbar}\left[\frac{1}{2}\mathbf{q}\cdot D_t\cdot B^{-1}_t\cdot \mathbf{q}+\frac{1}{2}\mathbf{q}'\cdot B^{-1}_t\cdot A_t\cdot \mathbf{q}'-\mathbf{q}'\cdot B^{-1}_t\cdot \mathbf{q}+\mathbf{q}\cdot\zeta(t)+\mathbf{q}'\cdot\eta(t)\right]}.
\end{eqnarray}
The equation (\ref{finalK}) is one of the main results of the present paper.
\subsection{Example 1:}
\noindent The Hamiltonian of describing a harmonic oscillator in $n$-dimensions under the influence of a classical vector field $f(t)=[f_1 (t),\cdots,f_n(t)]$ is
\begin{equation}\label{16}
\hat{H}=\frac{1}{2m}\hat{p}\cdot \hat{p}+\frac{1}{2}m\omega^2\hat{q}\cdot \hat{q}-f(t)\cdot\hat{q},
\end{equation}
where, $K_t=(1/m)\mathbf{1}_n$, $L_t=0$, $Z_t=m\omega^2\mathbf{1}_n$. In this case, $\dot{\vartheta}(t)=\gamma' \vartheta(t)$, and $\gamma'$ is a time-independent matrix defined by
\begin{equation}\label{17}
\gamma'=g\cdot \omega(t)=
\left(
\begin{array}{cc}
	 0 & \frac{1}{m}\mathbf{1}_n \\
	-m\omega^2 \mathbf{1}_n & 0 \\
\end{array}
\right).
\end{equation}
Therefore,
\begin{eqnarray}\label{18}
\vartheta(t) &=& e^{t\gamma'}\vartheta(0)=e^{t\gamma'}=
\left(
\begin{array}{cc}
	 \cos(\omega\,t)\mathbf{1}_n & \frac{\sin(\omega\,t)}{m\omega}\mathbf{1}_n \\
	-m\omega\sin(\omega\,t)\mathbf{1}_n & \cos(\omega\,t)\mathbf{1}_n
\end{array}
\right),\nonumber\\
\end{eqnarray}
and from Eq. (\ref{7-0}), we find $A_t=D_t=\cos(\omega\,t)\mathbf{1}_n$, $B_t=\mathbf{1}_n\sin(\omega\,t)/m\omega$ and $C_t=-m\omega\sin(\omega\,t)\mathbf{1}_n$. To find the vector fields $\eta(t)$ and $\xi(t)$, we use the Green function
\begin{eqnarray}\label{Green}
 G(t-t')&=&\theta(t-t')\vartheta(t)\vartheta^{-1}(t'),\nn\\
        &=& \theta(t-t')\left(
\begin{array}{cc}
	 \cos(\omega\,(t-t'))\mathbf{1}_n & \frac{\sin(\omega\,(t-t'))}{m\omega}\mathbf{1}_n \\
	-m\omega\sin(\omega\,(t-t'))\mathbf{1}_n & \cos(\omega\,(t-t'))\mathbf{1}_n
\end{array}
\right),
\end{eqnarray}
we have
\begin{eqnarray}
  \eta_i (t)&=&\int_0^t dt'\,\frac{\sin(\omega(t-t'))}{m\omega}\,f_i (t'),\,\,(i=1,\cdots,n),\nn \\
  \xi_i (t) &=& \int_0^t dt'\,\cos(\omega(t-t'))\,f_i (t').
\end{eqnarray}
Also, $\zeta(t)=\xi(t)-D_t\cdot B_t^{-1}\cdot\eta(t)$. Inserting these values into Eq. (\ref{finalK}), we find the propagator.
\section{Chain Of Harmonic Oscillators}
\noindent For a chain of linearly interacting harmonic oscillators with equal masses and couplings, the Hamiltonian in the presence of an external vector field $f(t)$ is
\begin{equation}\label{chain1}
 \hat{H}=\frac{1}{2}\hat{q}\cdot Z \cdot\hat{q}+\frac{1}{2}\hat{p}\cdot K\cdot\hat{p}-f(t)\cdot\hat{q},
\end{equation}
where $K=(1/m)\,\mathbf{1}_n$ and $Z$ is a time-independent $n\times n$ matrix that its form can be determined from the boundary conditions imposed on the chain. In this case, the matrix $\gamma=g\cdot \omega$ is time-independent, given by
\begin{equation}\label{per2}
\gamma=
\left(
 \begin{array}{cc}
 0 & \frac{1}{m} \mathbf{1}_n\\
 -Z & 0\\
 \end{array}
 \right).
\end{equation}
Now, from $\vartheta(t)=e^{t\,\gamma}\,\vartheta(0)=e^{t\,\gamma}$, we find
\begin{eqnarray}\label{per3}
\vartheta(t) &=&
\left(
\begin{array}{cc}
 A_t & B_t\\
C_t & D_t\\
 \end{array}
\right)\nn\\
&=&
\left(\begin{array}{cc}
 \cos(t\sqrt{Z/m}) & \frac{1}{\sqrt{mZ}}\,\sin(t\sqrt{Z/m})\\
 -\sqrt{mZ}\,\sin(t\sqrt{Z/m}) & \cos(t\sqrt{Z/m})\\
 \end{array}
 \right),
\end{eqnarray}
Therefore, $A_t=D_t=\cos(t\sqrt{Z/m})$, $B_t=\frac{1}{\sqrt{mZ}}\,\sin(t\sqrt{Z/m})$, and $C_t=-\sqrt{mZ}\,\sin(t\sqrt{Z/m})$. Also,
\begin{eqnarray}
  \eta_i (t)&=& \sum_{j=1}^n\int_0^t dt'\,\left(\frac{\sin(\sqrt{\frac{Z}{m}}(t-t'))}{\sqrt{mZ}}\right)_{ij}\,f_j (t'),\,\,(i=1,\cdots,n),\nn \\
  \xi_i (t) &=& \sum_{j=1}^n\int_0^t dt'\,\left(\cos(\sqrt{\frac{Z}{m}}(t-t'))\right)_{ij}\,f_j (t'),\nn\\
  \zeta_i(t)&=&\xi_i(t)-\sum_{j=1}^n (D_t\cdot B_t^{-1})_{ij}\eta_j(t).
\end{eqnarray}
Finally, the quantum propagator for the chain is
\begin{eqnarray}\label{per4}
&& K(\mathbf{q},t|\mathbf{q}',0)= \frac{1}{\sqrt{(2\pi i\hbar)^n \big|\det (\frac{\sin(t\sqrt{Z/m})}{\sqrt{m Z}})\big|}}\,e^{-\frac{i}{2m\hbar}\int_0^t dt'\,|\zeta(t')|^2}
\,e^{\frac{i}{\hbar}[\mathbf{q}\cdot\zeta(t)+\mathbf{q}'\cdot\eta(t)]}\nn\\
&& \times\,e^{\frac{i}{2\hbar}\left[\mathbf{q}\cdot\left\{\frac{\sqrt{mZ}\cos(t\sqrt{Z/m})}{\sin(t\sqrt{Z/m})}\right\}\cdot \mathbf{q}+\mathbf{q}'\cdot\left\{\frac{\sqrt{mZ}\cos(t\sqrt{Z/m})}{\sin(t\sqrt{Z/m})}\right\}\cdot \mathbf{q}'-2\mathbf{q}'\cdot\left\{\frac{\sqrt{mZ}}{\sin(t\sqrt{Z/m})}\right\}\cdot \mathbf{q}
\right]}.\nn\\
\end{eqnarray}
To proceed, let $\mathbf{v}_i$ be an eigenvector of the matrix $Z$ with eigenvalue $z_i$, ($Z\cdot \mathbf{v}_i=z_i\,\mathbf{v}_i,\,\,i=1,\cdots,n$), then the propagator Eq. (\ref{per4}) can be rewritten  in terms of the eigenvectors and eigenvalues of $Z$ as follows
\begin{eqnarray}\label{per4}
K(\mathbf{q},t|\mathbf{q}',0) &=& \sqrt{\frac{\prod\limits_{k=1}^n g(z_k)}{(2\pi i\hbar)^n}}\,e^{\frac{i}{2\hbar}\sum\limits_{k=1}^n \Big[\left[(\mathbf{q}\cdot \mathbf{v}_k)^2+(\mathbf{q}'\cdot \mathbf{v}_k)^2\right]\,f(z_k)-2(\mathbf{q}\cdot \mathbf{v}_k)(\mathbf{q}'\cdot \mathbf{v}_k)\,g(z_k)\Big]},\nn\\
&& \times \,e^{-\frac{i}{2m\hbar}\int_0^t dt'\,|\zeta(t')|^2}\,e^{\frac{i}{\hbar}[\mathbf{q}\cdot\zeta(t)+\mathbf{q}'\cdot\eta(t)]}.
\end{eqnarray}
where
\begin{eqnarray}\label{47}
f(z_k)=&& \sqrt{m z_k}\,\cot(t\sqrt{z_k/m}),\nn\\
g(z_k)=&& \frac{\sqrt{m z_k}}{\sin(t\sqrt{z_k/m})}.
\end{eqnarray}
By using the eigenvectors $\mathbf{v}_i$ we define the transformation matrix $V=[v_{ij}]$, where the ith column of the matrix $V$ is the eigenvector $\mathbf{v}_i$
\begin{equation}\label{Vmatrix}
V=
\left(
\begin{array}{cccc}
	 \mathbf{v}_{1} & \mathbf{v}_{2} & \cdots & \mathbf{v}_{n} \\
	\end{array}
\right).
\end{equation}
In the basis of the orthonormal eigenvectors $\mathbf{v}_i$, the position vector $\mathbf{q}$ has the components
\begin{equation}\label{newcor}
  \tilde{q}_i=\sum_{j=1}^n v_{ij}\,q_j=\mathbf{v}_i\cdot \mathbf{q},
\end{equation}
where $v_{ij}$ is the jth component of the eigenvector $\mathbf{v}_i$, and similarly, we define $\tilde{q}'_i=\mathbf{v}_i\cdot \mathbf{q}'$, $\tilde{\zeta}(t)_i=\mathbf{v}_i\cdot\zeta(t)$,
$\tilde{\eta}(t)_i=\mathbf{v}_i\cdot\eta(t)$. Therefore, the propagator Eq. (\ref{per4}) can be rewritten in new coordinates $\tilde{\mathbf{q}}$ and $\tilde{\mathbf{q}}'$ as
\begin{eqnarray}\label{propagator}
K(\tilde{\mathbf{q}},t|\tilde{\mathbf{q}}',0) &=& \sqrt{\frac{\prod\limits_{k=1}^n g(z_k)}{(2\pi i\hbar)^n}}\, e^{-\frac{i}{2m\hbar}\int_0^t dt'\,|\zeta(t')|^2}\nn\\
&& \times\,e^{\frac{i}{\hbar}\sum\limits_{k=1}^n \Big[\left[\frac{1}{2}(\tilde{q}_k)^2+\frac{1}{2}(\tilde{q}'_k)^2\right]\,f(z_k)-\tilde{q}_k\,\tilde{q}'_k\,g(z_k)+\tilde{q}_k\,\tilde{\zeta}(t)_k+\tilde{q}'_k\,\tilde{\eta}(t)_k\Big]}.
\end{eqnarray}
or equivalently
\begin{eqnarray}\label{Eqpropagator}
K(\tilde{\mathbf{q}},t|\tilde{\mathbf{q}}',0) &=& \sqrt{\prod\limits_{k=1}^n \left(\frac{m\omega_k}{2\pi i\hbar\sin(\omega_k t)}\right)}\,\, e^{-\frac{i}{2m\hbar}\int_0^t dt'\,|\zeta(t')|^2}\nn\\
&& \times\,e^{\frac{im\omega_k}{2\hbar\sin(\omega_k t)}\sum\limits_{k=1}^n \Big[\left[(\tilde{q}_k^2+\tilde{q}'_k)^2)\right]
\cos(\omega_k t)-\tilde{q}_k\,\tilde{q}'_k+2\tilde{q}_k\,\tilde{\zeta}(t)_k+2\tilde{q}'_k\,\tilde{\eta}(t)_k\Big]},
\end{eqnarray}
where we made use of Eqs. (\ref{47}) and defined the eigenfrequencies $\omega_k=\sqrt{z_k/m}$. Note that the propagator Eq. (\ref{propagator}) is the product of the propagators of $n$ driven harmonic oscillator as we expected from the diagonalization of the Hamiltonian in terms of the normal modes. To find the time-evolution of an arbitrary initial state $|\psi(0)\ra$, it should be represented in the new coordinates $\la \tilde{\mathbf{q}}|\psi(0)\ra=\psi(\tilde{\mathbf{q}},0)$, then
\begin{equation}\label{evolution}
 \psi(\tilde{\mathbf{q}},t)=\int d\tilde{\mathbf{q}}'\,K(\tilde{\mathbf{q}},t|\tilde{\mathbf{q}}',0)\,\psi(\tilde{\mathbf{q}}',0).
\end{equation}
Note that since the transformation matrix is orthogonal $V\,V^T=\mathbf{1}$, the Jacobian determinant is unity and accordingly $d\tilde{\mathbf{q}}'=d\mathbf{q}'$.
\subsection{Periodic boundary conditions}
%
\begin{figure}
\centering
\includegraphics[scale=0.5]
{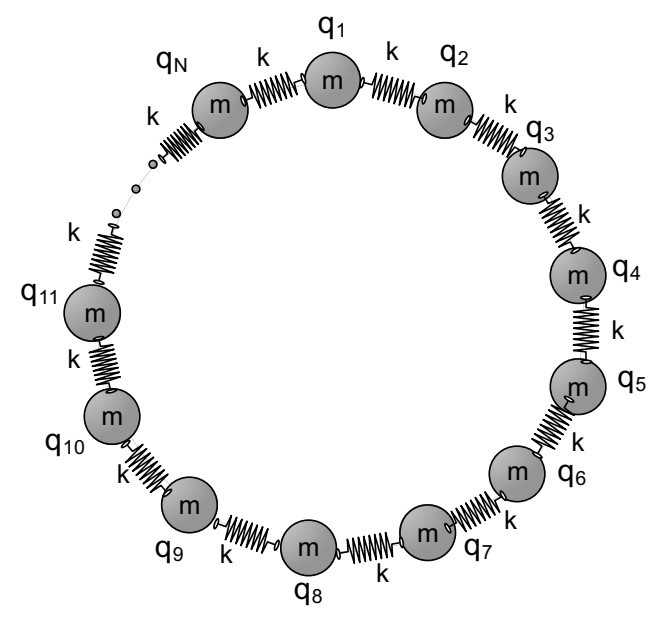}
\caption{The Chain Of linearly interacting Harmonic Oscillators with periodic boundary condition $\hat{q}_{N+1}=\hat{q}_1$.}
\end{figure}
\noindent For a linear chain with periodic boundary condition (see Fig. 1), we have $q_1=q_{N+1}$, and the Hamiltonian is
\begin{equation}\label{per1}
\hat{H}=\sum_{i=1}^{N}\frac{1}{2m}\hat{p}_i^2+\sum_{i=1}^{N}\frac{1}{2}m\omega_0^2\Big(\hat{q}_{i+1} - \hat{q}_i\Big)^2,
\end{equation}
where the matrices $Z$ and $K$ are defined by
\begin{eqnarray}
&& Z_{ij}=2m\omega_0^2\,\delta_{i,j}-m\omega_0^2\,(\delta_{i,j+1}+\delta_{i,j-1}),\,\,\,(i,j=1,\cdots,n),\nn\\
&& K=(1/m)\,\mathbf{1}_n.
\end{eqnarray}
%
\subsection{Example 2:}
%
\begin{figure}\label{Fig2}
\centering
\includegraphics[scale=0.3]
{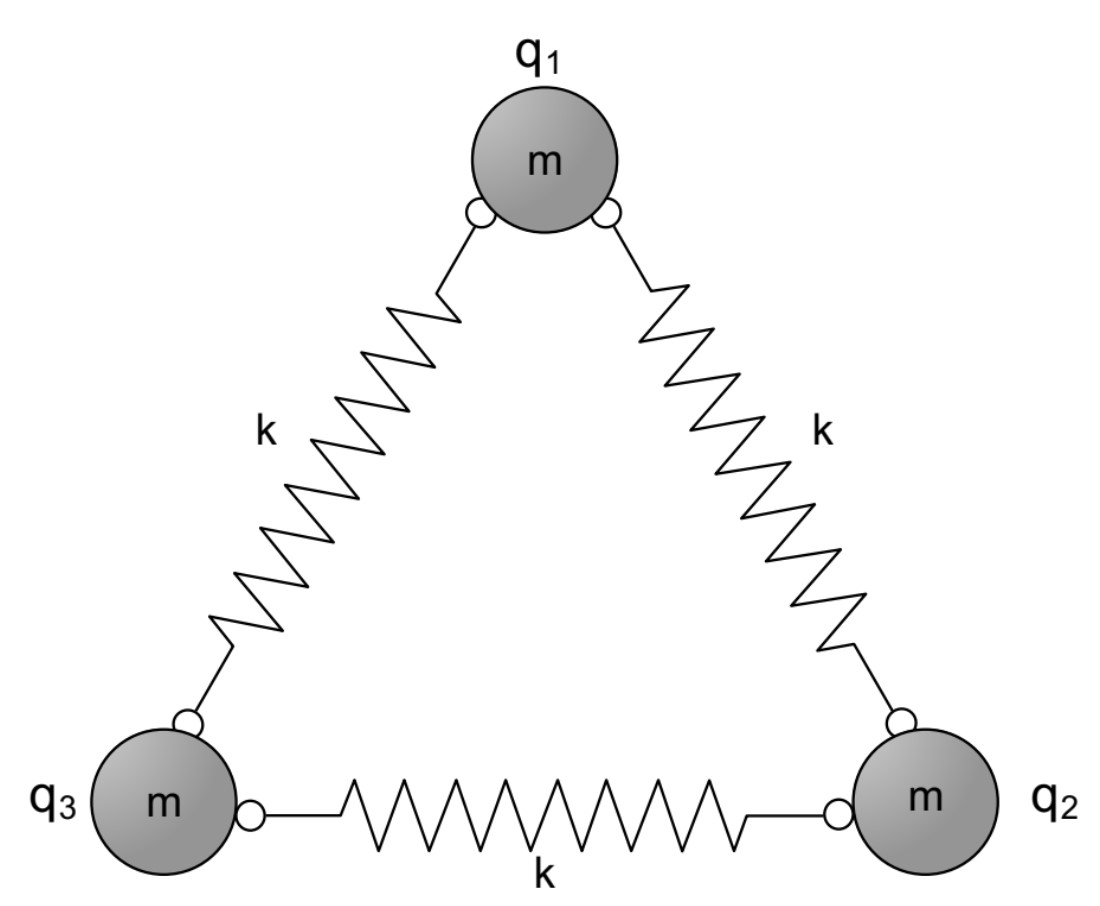}
\caption{Three identical coupled oscillators with periodic condition $\hat{q}_1=\hat{q}_4$.}
\end{figure}
\noindent For three linearly interacting oscillators (see Fig. 2), we set $N=3$, and obtain
\begin{equation}\label{H3p}
 \hat{H}=\sum_{i=1}^3 \frac{\hat{p}^2_i}{2m}+\frac{1}{2}m\omega^2_0\,\left[(q_1-q_2)^2+(q_2-q_3)^2+(q_1-q_3)^2\right].
\end{equation}
\begin{equation}\label{28}
K=
\left(
\begin{array}{ccc}
	 1/m & 0 & 0 \\
	0 & 1/m & 0 \\
    0 & 0 & 1/m
\end{array}
\right),
Z=
\left(
\begin{array}{ccc}
	2m\omega^2_0 & -m\omega^2_0 & -m\omega^2_0 \\
	-m\omega^2_0 & 2m\omega^2_0 & -m\omega^2_0 \\
    -m\omega^2_0 & -m\omega^2_0 & 2m\omega^2_0
\end{array}
\right).
\end{equation}
The eigenvectors and eigenvalues of the matrix $Z$ are
\begin{equation}
z_1=3m\omega^2_0\rightarrow \mathbf{v}_1=
\left(
\begin{array}{c}
	-1/\sqrt{2} \\
	0 \\
    1/\sqrt{2}
\end{array}
\right),
\end{equation}
\begin{equation}
 z_2=3m\omega^2_0\rightarrow \mathbf{v}_2=
\left(
 \begin{array}{c}
	 1/\sqrt{6} \\
	-2/\sqrt{6} \\
     1/\sqrt{6}  \\
\end{array}
\right),
\end{equation}
\begin{equation}\label{Vmatrix}
z_3=0\rightarrow \mathbf{v}_3=
\left(
\begin{array}{c}
	1/\sqrt{3} \\
	1/\sqrt{3} \\
    1/\sqrt{3}
\end{array}
\right).
\end{equation}
Therefore, the transformation matrix is
\begin{equation}
  V=
\left(
\begin{array}{ccc}
  -1/\sqrt{2} & 1/\sqrt{6} & 1/\sqrt{3} \\
	0 & -2/\sqrt{6} & 1/\sqrt{3} \\
    1/\sqrt{2} & 1/\sqrt{6} &  1/\sqrt{3}
    \end{array}
\right).
\end{equation}
The corresponding eigenfrequencies are $\omega_1=\omega_2=\sqrt{3}\,\omega_0$, and $\omega_3=0$ also $f(z_1)=f(z_2)=\sqrt{3}\,m_0\cot(\sqrt{3}\,\omega_0 t)$, $f(z_3)=m/t$, $g(z_1)=g(z_2)=\sqrt{3}\,m\omega_0/\sin(\sqrt{3}\,\omega_0 t)$ and $g(z_3) = m/t$.
%
\subsection{Dirichlet boundary condition}
%
\begin{figure}\label{Fig3}
\centering
\includegraphics[scale=0.4]
{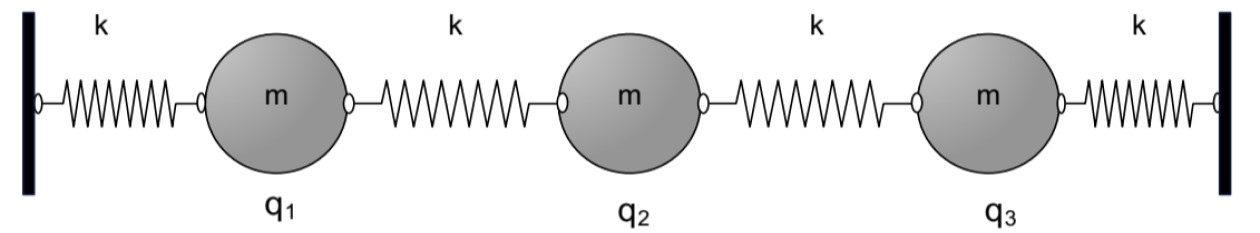}
\caption{Three interacting oscillators with the Dirichlet boundary conditions.}
\end{figure}
\noindent For a Linear chain of harmonic oscillators with Dirichlet boundary condition $q_0 = q_{N+1}=0$, The total Hamiltonian for N-oscillators is
\begin{equation}\label{51l}
\hat{H}=\sum_{i=1}^{N}\frac{1}{2m}\hat{p}_i^2+\sum_{i=1}^{N}\frac{1}{2}m\omega_0^2\Big(\hat{q}_{i+1} - \hat{q}_i\Big)^2.
\end{equation}
\subsection{Example 3:}
For three harmonic oscillators under the influence of an external force field $f(t)=[f_1 (t),f_2 (t),f_3(t)]^T$, (see Fig. 3), we have
\begin{equation}\label{28}
K=
\left(
\begin{array}{ccc}
	 1/m & 0 & 0 \\
	0 & 1/m & 0 \\
    0 & 0 & 1/m
\end{array}\right), Z=
\left(
\begin{array}{ccc}
	2m\omega^2_0 & -m\omega^2_0 & 0 \\
	-m\omega^2_0 & 2m\omega^2_0 & -m\omega^2_0 \\
    0 & -m\omega^2_0 & 2m\omega^2_0
\end{array}
\right).
\end{equation}
The eigenvectors and eigenvalues of the matrix $Z$ are
\begin{eqnarray}\label{54l}
&& z_1=(2+\sqrt{2})\,m\omega_0^2\rightarrow \mathbf{v}_1=
\left(
\begin{array}{c}
	1/2 \\
	-\sqrt{2}/2 \\
    1/2
\end{array}
\right),\nn\\
&& z_2=2\,m\omega_0^2\rightarrow \mathbf{v}_2=
\left(
\begin{array}{c}
	1/\sqrt{2} \\
	0 \\
    -1/\sqrt{2}
\end{array}
\right),\nn\\
&& z_3=(2-\sqrt{2})\,m\omega_0^2\rightarrow \mathbf{v}_3=
\left(
\begin{array}{c}
	1/2 \\
	\sqrt{2}/2 \\
    1/2
\end{array}
\right),
\end{eqnarray}
and the transformation matrix is
\begin{equation}
V=
\left(
\begin{array}{ccc}
  1/2 & \sqrt{2}/2 & 1/2 \\
	-\sqrt{2}/2 & 0 & \sqrt{2}/2 \\
    1/2 & -\sqrt{2}/2 & 1/2
  \end{array}
\right).
\end{equation}
The corresponding eigenfrequencies are $\omega_1=\sqrt{2+\sqrt{2}}\,\omega_0$, $\omega_2=\sqrt{2}\,\omega_0$, and $\omega_3=\sqrt{2-\sqrt{2}}\,\omega_0$ also
\begin{eqnarray}
&&  f(z_1)=\sqrt{2+\sqrt{2}}\,m\omega_0\,\cot(\sqrt{2+\sqrt{2}}\omega_0 t),\nn\\
&&  f(z_2)=\sqrt{2}\,m\omega_0\,\cot(\sqrt{2}\omega_0 t),\nn \\
&&  f(z_3)=\sqrt{2-\sqrt{2}}\,m\omega_0\,\cot(\sqrt{2-\sqrt{2}}\omega_0 t),\nn\\
&&  g(z_1)= \frac{\sqrt{2+\sqrt{2}}\,m\omega_0}{\sin(\sqrt{2+\sqrt{2}}\,\omega_0 t)},\nn \\
&&  g(z_2)= \frac{\sqrt{2}\,m\omega_0}{\sin(\sqrt{2}\,\omega_0 t)},\nn\\
&&  g(z_3)= \frac{\sqrt{2-\sqrt{2}}\,m\omega_0}{\sin(\sqrt{2-\sqrt{2}}\,\omega_0 t)}.
\end{eqnarray}
Also,
\begin{eqnarray}
  \eta_i (t) &=& \sum_{j=1}^3 \int_0^t dt'\,\frac{\sin \omega_j (t-t')}{m\omega_j}\,f_j (t'), \,\, i=1,2,3,\nn \\
  \xi_i (t)  &=& \sum_{j=1}^3 \int_0^t dt'\,\cos(m\omega_j (t-t')\,f_j (t'),
\end{eqnarray}
and the exact propagator can be obtained using Eq. (\ref{propagator}) or Eq. (\ref{Eqpropagator}).
\section{Interacting oscillators in ladder operators representation}
\subsection{Propagation of excitation along the oscillator chain}
\noindent
In this section, in the framework of ladder operators, we study the state propagation along a chain of $n$ linearly interacting oscillators (Fig. 4). The oscillators are considered to be identical with frequency $\omega_0$ and the coupling constant is $g$, \cite{shareef2022quantum}. The Hamiltonian is
\begin{equation}\label{chain1}
  \hat{H}(t)=\sum_{k=1}^{n}\hbar \omega_0 \hat{a}_{k}^\dag \hat{a}_{k}+\sum_{k=1}^{n-1}\hbar g\,(\hat{a}_{k} \hat{a}_{k+1}^\dag + \hat{a}_{k}^\dag \hat{a}_{k+1}).
\end{equation}
The Hamiltonian in the matrix form is
\begin{equation}\label{hchain}
\hat{H}(t)=
\left(
\begin{array}{cccc}
	\hat{a}^{\dag}_1 & \hat{a}^{\dag}_2 &...& \hat{a}^{\dag}_n\\
\end{array}
\right)\cdot\Lambda\cdot\left(
\begin{array}{c}
	\hat{a}_1 \\
    \hat{a}_2 \\
    \vdots\\
	\hat{a}_n
\end{array}
\right),
\end{equation}
where the coefficient matrix $\Lambda$ is defined by
\begin{equation}
\Lambda=
\left(
\begin{array}{cccccc}
	 \hbar\omega_0 & \hbar g & 0 & 0 & \cdots & 0  \\
	 \hbar g & \hbar\omega_0 & \hbar g & 0 & \cdots & 0 \\
      0  & \hbar g & \hbar\omega_0 & \hbar g & \dots & 0 \\
      \vdots    & \cdots     & \ddots  & \ddots & \ddots & \vdots \\
      0 & \cdots & 0 & \hbar g & \hbar\omega_0 & \hbar g \\
      0  & \cdots & 0 & 0 & \hbar g & \hbar\omega_0\\
\end{array}
\right).
\end{equation}
The coefficient matrix $\Lambda$ is a real symmetric matrix that can be diagonalized by the orthogonal matrix $T$
\begin{equation}\label{chain3}
  T^{t}\,\Lambda\,T=D,
\end{equation}
where $D=diag(\hbar\lambda_1,\hbar\lambda_2,\cdots,\hbar\lambda_n)$ is a diagonal matrix with eigenvalues of $\Lambda$ on its diameter. Let us define the new operators $\hat{A}_k$ by
\begin{equation}\label{chain4}
  \hat{A}_{i}=\sum_{j=1}^nT_{ji}\,\hat{a}_j,
\end{equation}
then the Hamiltonian Eq. (\ref{hchain}) can be written as
\begin{equation}\label{chain5}
  \hat{H}(t)=\sum_{k=1}^{m}\hbar\lambda_k\, \hat{A}_{k}^\dag \hat{A}_{k}.
\end{equation}
\begin{figure}[t]
  \centering
  \includegraphics[width=.7\textwidth]{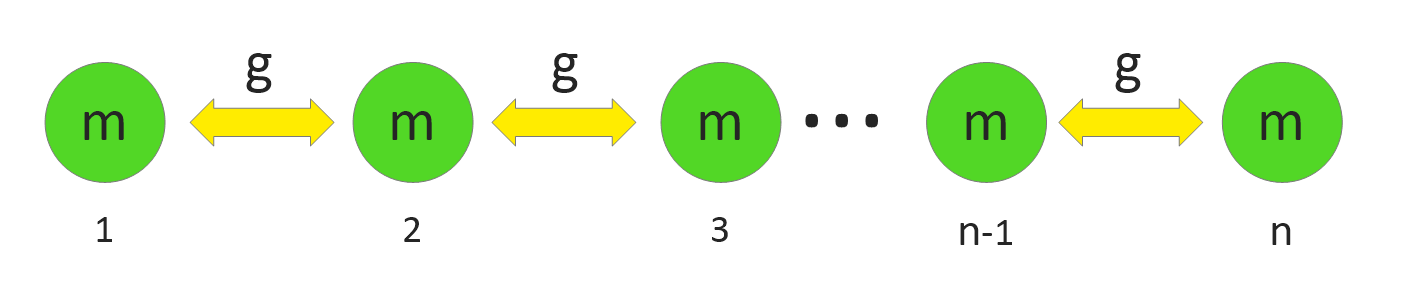}
  \caption{A chain of identical oscillators interacting trough a constant coupling $g$.}
  \label{chainfig}
\end{figure}
The new operators $\hat{A}_{k}$ and $ \hat{A}_{k}^\dag$ evolve in Heisenberg picture as $\hat{A}_{k}(t)=e^{-it\lambda_k}\,\hat{A}_{k}$ and $\hat{A}_{k}^\dag(t)=e^{it\lambda_k}\,\hat{A}_{k}^\dag$. The new frequencies (eigenvalues) $\lambda_k$ and the $j$th component of the eigenvector $|\mathbf{V}^k\ra$, are given by
\begin{eqnarray}\label{chain6}
  \lambda_k &=& \omega+2g\,\cos\left(\frac{k\pi}{k+1}\right),\nn \\
  V^k_j &=& \sqrt{\frac{2}{n+1}}\sin\left(\frac{kj\pi}{n+1}\right),
\end{eqnarray}
respectively. Now the Hamiltonian is decomposed into independent Hamiltonians and the evolution operator is
\begin{equation}\label{chain7}
  \Uh(t)=\prod_{k=1}^n\,\otimes e^{-i\lambda_k t\,\hat{A}^\dag_k \hat{A}_k}.
\end{equation}
\subsection{Propagation of a coherent state}
Let the initial state of the chain be the separable state
\begin{eqnarray}\label{chain8}
  \hat{\rho} (0) &=& |\alpha\ra_{11}\la \alpha|\otimes|0\ra_{22}\la 0|\otimes\cdots\otimes|0\ra_{nn}\la 0|,\nn\\
                 &=& e^{\alpha\hat{a}^\dag_1-\bar{\alpha}\hat{a}_1}\,|0\ra\la 0|\,e^{-\alpha\hat{a}^\dag_1+\alpha\hat{a}_1},
\end{eqnarray}
where at $t=0$ the first oscillator is prepared in a coherent state and the remaining oscillators are in their ground state. By inserting Eqs. (\ref{chain4}) into Eq. (\ref{chain8}), we obtain
\begin{eqnarray}\label{chain9}
  \hat{\rho} (0) &=& e^{\alpha \sum_j T_{ij}\hat{A}^\dag_j-\bar{\alpha} \sum_j T_{ij}\hat{A}_j}\,|0\ra\la 0|\,e^{-\alpha T_{ij}\hat{A}^\dag_j+\bar{\alpha} T_{ij}\hat{A}_j},\nn\\
   &=& \prod_{j=1}^n e^{\alpha T_{ij}\hat{A}^\dag_j-\bar{\alpha} T_{ij}\hat{A}_j}\,|0\ra\la 0|\,e^{-\alpha T_{ij}\hat{A}^\dag_j+\bar{\alpha} T_{ij}\hat{A}_j},\nn\\
   &=& |\alpha T_{11}\ra_{11}\la\alpha T_{11}|\otimes\cdots\otimes|\alpha T_{1n}\ra_{nn}\la \alpha T_{1n}|.
\end{eqnarray}
Therefore,
\begin{eqnarray}\label{chain10}
  \hat{\rho} (t)  &=& |\alpha T_{11} e^{-i\lambda_1 t}\ra_{11}\la\alpha T_{11} e^{-i\lambda_1 t}|\otimes\cdots\otimes|\alpha T_{1n} e^{-i\lambda_n t}\ra_{nn}\la \alpha T_{1n} e^{-i\lambda_n t}|,\nn\\
  &=& \prod_{j=1}^n e^{(\alpha T_{1j} e^{-i\lambda_j t})\hat{A}^\dag_j-(\bar{\alpha} T_{1j} e^{i\lambda_j t})\hat{A}_j}\,|0\ra\la 0|\,\prod_{j=1}^n e^{(-\alpha T_{1j} e^{-i\lambda_j t})\hat{A}^\dag_j+(\bar{\alpha} T_{1j} e^{i\lambda_j t})\hat{A}_j},\nn\\
  &=& |\sum_{j}\alpha T_{1j} e^{-i\lambda_j t}T_{1j}\ra_{11}\la \sum_{j}\alpha T_{1j} e^{-i\lambda_j t}T_{1j}|\otimes\cdots\nn\\
  && \otimes|\sum_{j}\alpha T_{1j} e^{-i\lambda_j t}T_{1j}\ra_{nn}\la \sum_{j}\alpha T_{nj} e^{-i\lambda_j t}T_{nj}|.
\end{eqnarray}
On the other hand
\begin{equation}
  e^{-i t D/\hbar}=
\left(
\begin{array}{ccc}
	             e^{-i\lambda_1 t} & \cdots & 0\\
                 \vdots & \ddots & \vdots\\
                 0 & \cdots &  e^{-i\lambda_1 t}
\end{array}
\right)
=e^{-\frac{i}{\hbar} t T^t\,\Lambda\,T}=T^t\,e^{-\frac{i}{\hbar}t\Lambda}\,T,
\end{equation}
therefore,
\begin{equation}\label{chain11}
 T\,e^{-\frac{i}{\hbar}t\,D}\,T^t=e^{-\frac{i}{\hbar} t\, \Lambda}.
\end{equation}
Now using Eq. (\ref{chain11}), we obtain
\begin{equation}\label{chain12}
  (T\,e^{-\frac{i}{\hbar}t\,D}\,T^t)_{1l}=\sum_{j,k} T_{1j}\,e^{-it\lambda_j}\,\delta_{jk}\,T_{lk}=(e^{-\frac{i}{\hbar} t \,\Lambda})_{1l},
\end{equation}
leading to the final result
\begin{eqnarray}
  \hat{\rho}(t) &=& |\alpha (e^{-\frac{i}{\hbar}t\,\Lambda})_{11}\ra_{11}\la \alpha (e^{-\frac{i}{\hbar}t\,\Lambda})_{11}|\otimes\cdots\otimes|\alpha (e^{-\frac{i}{\hbar}t\,\Lambda})_{1n}\ra_{nn}\la \alpha (e^{-\frac{i}{\hbar}t\,\Lambda})_{1n}|.\nn\\
\end{eqnarray}
Having the density matrix, we can find the mean number of excitations for the jth oscillator as
\begin{eqnarray}\label{chain12}
  \la \hat{a}^\dag_j \hat{a}_j\ra_t &=& tr[\hat{\rho} (t)\,\hat{a}^\dag_j \hat{a}_j],\nn\\
                                    &=& \la \alpha(e^{-\frac{i}{\hbar}t\,\Lambda})_{1j}|\hat{a}^\dag_j \hat{a}_j|\alpha(e^{-\frac{i}{\hbar}t\,\Lambda})_{1j}\ra,\nn\\
                                    &=& |\alpha|^2\,|(e^{-\frac{i}{\hbar}t\,\Lambda})_{1j}|^2.
\end{eqnarray}
Now we have
\begin{eqnarray}
  (e^{-\frac{i}{\hbar}t\,\Lambda})_{1j} &=& \la 1|e^{-\frac{i}{\hbar}t\,\Lambda}|j\ra=\left(\sum_{k=1}^n\,e^{-it\,\lambda_k}\,|\mathbf{V}^k\ra\la \mathbf{V}^k|\right)_{1j},\nn \\
                          &=& \sum_{k=1}^n\,e^{-it\,\lambda_k}\,\la 1|\mathbf{V}^k\ra\la \mathbf{V}^k|j\ra,\nn\\
                          &=& \left(\frac{2}{n+1}\right)\,\sum_{k=1}^n\,e^{-it\,(\omega+2g\cos(\frac{k\pi}{n+1}))}\,\sin\left(\frac{k\pi}{n+1}\right)\sin\left(\frac{kj\pi}{n+1}\right),\nn\\
\end{eqnarray}
where the basis $\{|j\ra\}_{j=1}^n$ are defined by
\begin{equation}\label{chain13}
  |1\ra=
\left(
\begin{array}{c}
	            1\\
                0\\
                \vdots\\
                0
\end{array}
\right),\,\,\,|2\ra=
\left(
\begin{array}{c}
	                               0\\
                                   1\\
                              \vdots\\
                                   0
\end{array}
\right),\cdots,|n\ra=
\left(
\begin{array}{c}
	                                                                   0\\
                                                                       0\\
                                                                   \vdots\\
                                                                       1
\end{array}
\right).
\end{equation}
Note that $\la 1|e^{-\frac{i}{\hbar}t\,\Lambda}|j\ra$ is like a probability amplitude for transition from the initial state $|j\ra$ to the final state $|1\ra$ at time $t$. This kernel or propagator is corresponding to the Hamiltonian $\Lambda$ in an $n$-dimensional Hilbert space spanned by the standard basis $\{|j\ra\}_{j=1}^n$. Therefore, the probability of a transition from $|j\ra$ to $|i\ra$ during the time $t$ is
\begin{eqnarray}\label{chain14}
&&  P_{j\,i} (t)=| (e^{-\frac{i}{\hbar}t\,\Lambda})_{ij}|^2,\nn\\
&& =\left(\frac{2}{n+1}\right)^2\,\left|\sum_{k=1}^n\,e^{-it\,(\omega+2g\cos(\frac{k\pi}{n+1}))}
  \,\sin\left(\frac{ik\pi}{n+1}\right)\sin\left(\frac{kj\pi}{n+1}\right)\right|^2.
\end{eqnarray}
In particular, for $j=1$ and $i=n$, we find
\begin{eqnarray}\label{chain15}
 \frac{ \la \hat{a}^\dag_n \hat{a}_n\ra_t}{ \la \hat{a}^\dag_1 \hat{a}_1\ra_0} &=& P_{n\,1} (t),\nn \\
   &=& \left(\frac{2}{n+1}\right)^2\,\left|\sum_{k=1}^n\,(-1)^{k+1}\,e^{-2igt\cos(\frac{k\pi}{n+1})}
  \,\sin^2\left(\frac{k\pi}{n+1}\right)\right|^2.
\end{eqnarray}
\subsection{The propagation of an arbitrary state $|\psi\ra_1$}
\noindent Let the initial state of the chain be the separable state
\begin{eqnarray}\label{chain16}
  |\Psi(0)\ra &=& |\psi\ra_1\otimes|0\ra_2\otimes\cdots\otimes|0\ra_n,\nn\\
              &=& |\psi\ra_1\,|0\ra_2\,\cdots\,|0\ra_n,
\end{eqnarray}
where now the first oscillator is prepared in an arbitrary state $|\psi\ra_1$ which can be expanded in the number state basis $\{|n\ra_1\}$ belonging to the first oscillator Hilbert space
\begin{eqnarray}\label{chain17}
&&  |\psi\ra_1 = \sum_{k=0}^\infty c_k\,|k\ra_1,\nn\\
&&  |k\ra_1=\frac{(\hat{a}_1^{\dag k})}{\sqrt{k!}}\,|0\ra_1.
\end{eqnarray}
Now from Eq. (\ref{chain4}), we have $\hatad_1=\sum\limits_l T_{1l}\hatAd_l$, and we can rewrite $|\Psi(0)\ra$ as
\begin{eqnarray}\label{chain18}
   |\Psi(0)\ra &=& \sum_{k=0}^\infty \frac{c_k}{\sqrt{k!}}\,(T_{11}\,\hatAd_1+T_{12}\,\hatad_2+\cdots+T_{1n}\,\hatad_n)^k\,|0\ra,\nn \\
   &=&  \sum_{k=0}^\infty \frac{c_k}{\sqrt{k!}}\sum_{\sum\limits_{i=1}^n=k}\,C^k_{j_1,j_2,\cdots,j_n}
\,(T_{11}\hatAd_1)^{j_1}\cdots(T_{1n}\hatAd_n)^{j_n}\,|0\ra,\nn\\
\end{eqnarray}
where
\begin{equation}\label{binomial}
C^k_{j_1,j_2,\cdots,j_n}=\frac{k!}{j_1!\,j_2!\,\cdots j_n!}.
\end{equation}
Therefore, by using $\hat{A}_{k}(t)=e^{-it\lambda_k}\,\hat{A}_{k}$ and $\hat{A}_{k}^\dag(t)=e^{it\lambda_k}\,\hat{A}_{k}^\dag$ as well as noting that $\hat{U}^\dag (t)=\hat{U}(-t)$, the evolved state at time $t$ is
\begin{eqnarray}\label{chain19}
   |\Psi(t)\ra =\sum_{k=0}^\infty \frac{c_k}{\sqrt{k!}}\sum_{\sum_{i=1}^n j_i=k}
\,C^k_{j_1,j_2,\cdots,j_n}\,\prod_{k=1}^n\,(e^{-it\lambda_k}\,T_{1k}\hatAd_1)^{j_k}\,|0\ra.
\end{eqnarray}
By inserting $\hatAd_j=\sum\limits_l T_{lj}\,\hatad_l$ into Eq. (\ref{chain19}), and using
\begin{eqnarray}\label{chain20}
  \sum_{m=1}^n T_{1m}\,e^{-it\lambda_m}\,(T_{pm})\,\hatad_p &=& \sum_{p=1}^n\,(T\,e^{-\frac{i}{\hbar}t\,D}\,T^t)_{1p}\,\hatad,\nn \\
   &=& \sum_{p=1}^n\,(e^{-\frac{i}{\hbar}t\,\Lambda})_{1p}\,\hatad_p,
\end{eqnarray}
we will find
\begin{eqnarray}\label{chain21}
&& |\Psi(t)\ra = \sum_{k=0}^\infty \frac{c_k}{\sqrt{k!}}\,\left(\sum_{p=1}^n \,(e^{-\frac{i}{\hbar}t\,\Lambda})_{1p}\,\hatad_p\right)^k\,|0\ra,\nn\\
&& = \sum_{k=0}^\infty\sum_{\sum\limits_{i=1}^n k_i=k}\,\frac{c_k}{\sqrt{k!}}\,\sqrt{\prod\limits_{l=1}^n k_l !}
\,C^k_{k_1,k_2,\cdots,k_n}
\,\prod\limits_{l=1}^n\mu_l^{k_l}\,|k_1\ra_1\,|k_2\ra_2\,\cdots\,|k_n\ra_n,\nn\\
&& =\sum_{k=0}^\infty\,c_k\,\sum_{\sum\limits_{i=1}^n k_i=k}\,\sqrt{C^k_{k_1,k_2,\cdots,k_n}}\,\mu_1^{k_1}\,\mu_2^{k_2}\cdots\mu_n^{k_n}\,|k_1\ra_1\,|k_2\ra_2\,\cdots\,|k_n\ra_n,\nn\\
\end{eqnarray}
where for simplicity, we have defined $\mu_j=(e^{-\frac{i}{\hbar}t\,\Lambda})_{1j}$. Now let us calculate the mean excitation $\la \Psi(t)|\hatad_j\hatad_j|\Psi(t)\ra$. We have
\begin{eqnarray}\label{chain22}
  \la \Psi(t)|\hatad_j\hatad_j|\Psi(t)\ra &=& \sum_{k=0}^\infty |c_k|^2\,\sum_{\sum\limits_{i=1}^n k_i=k}
\,C^k_{k_1,k_2,\cdots,k_n}\,\mu_1^{k_1}\,\mu_2^{k_2}\cdots\mu_n^{k_n}\,k_j,\nn \\
  &=& \sum_{k=0}^\infty |c_k|^2\,(x_j\,\frac{\partial}{\partial x_j})\,(x_1+x_2+\cdots+x_j+\cdots+x_n)^k,\nn\\
  &=& {}_1\la \psi|\hatad_1\hata_1|\psi\ra_1\,|\mu_j|^2={}_1\la \psi|\hatad_1\hata_1|\psi\ra_1\,|\la 1|e^{-\frac{i}{\hbar}t\,\Lambda}|j\ra|^2.
\end{eqnarray}
In deriving Eq. (\ref{chain22}), we assumed $x_j=|\mu_j|^2$, and made use of the following identities
\begin{equation}\label{chain23}
  \sum_{k=1}^n x_k=\sum_{k=1}^n |\mu_k|^2=\sum_{k=1}^n\,(e^{-\frac{i}{\hbar}t\,\Lambda})_{1k}(e^{\frac{i}{\hbar}t\,\Lambda})_{k1}=1.
\end{equation}
From Eq. (\ref{chain22}),  we deduce that the ratio of the mean excitations for the $j$th oscillator and the initial oscillator, is independent on the initial state of the first oscillator given by
\begin{equation}\label{chain24}
  \frac{ \la \Psi(t)|\hatad_j\hatad_j|\Psi(t)\ra}{{}_1\la \psi|\hatad_1\hata_1|\psi\ra_1}=|\la 1|e^{-\frac{i}{\hbar}t\,\Lambda}|j\ra|^2.
\end{equation}
Note that Eq. (\ref{chain24}) is the probability of transition from the state $|1\ra$ to the state $|j\ra$ in the Hilbert space of a system described by the Hamiltonian $\Lambda$ acting on a $n$-dimensional Hilbert space spanned by the basis defined in Eq. (\ref{chain13}).
\subsection{Speed of excitation propagation along the chain}
\noindent Let the first oscillator be prepared in an arbitrary state $|\psi\ra_1$, this state evolves and the signal reaches the end oscillator ($n$th oscillator) at the scaled time $\tau=g t$, and disturbs the $n$th oscillator around its ground state. In Fig. 5, the probability amplitude $P(n,\tau)=|\la 1|e^{-\frac{i}{\hbar}t\,\Lambda}|n\ra|^2$ is depicted in terms of the number of oscillators $n$, and the scaled timed $\tau$. From Fig. 5, it seems that the first signal reaches the $n$th oscillator at the scaled time $\tau$ along a straight line. To elaborate more on this, we depicted the first maxima of the probability amplitude $P(n,\tau)$ projected on the $\tau-n$ plane in Fig. 6, which is a straight line showing the fact that the speed of propagation is independent on $n$, as expected.
\begin{figure}
\centering
\includegraphics[scale=0.6]{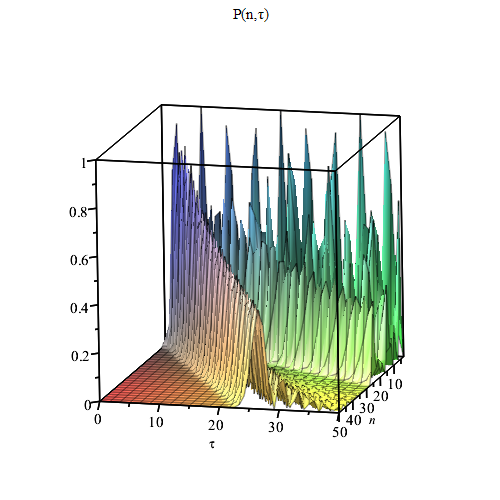}
\caption{(Color online) The probability amplitude $P(n,\tau)$ in terms of the number of oscillators $n$, and the scaled timed $\tau$. It is seen that the first signal reaches the $n$th oscillator at the scaled time $\tau$ along a straight line (see Fig. 6).}
\end{figure}
%
\begin{figure}
\centering
\includegraphics[scale=0.4]{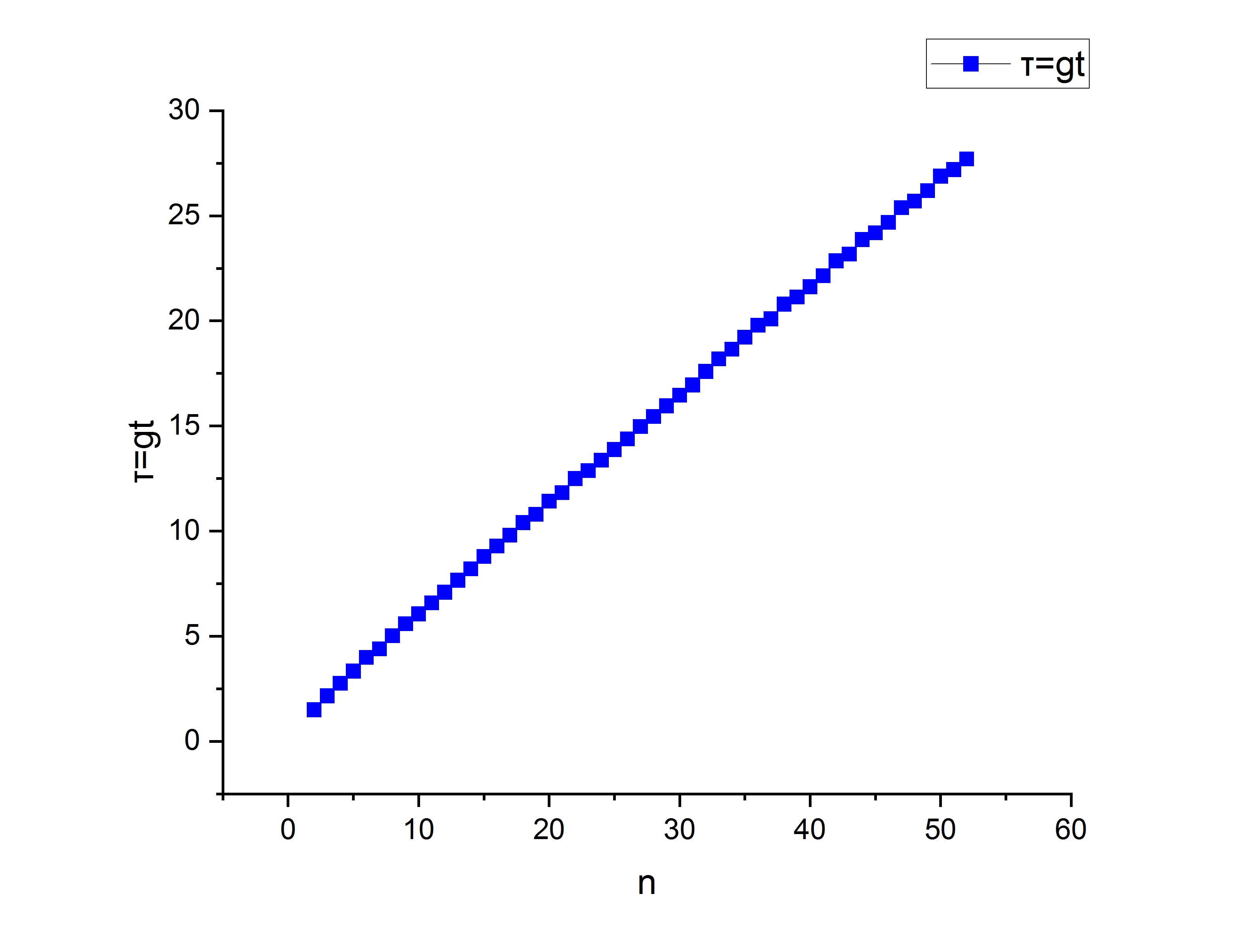}
\caption{(Color online) The location of the first maxima of the function $P(n,\tau)$ illustrated in Fig. 2, projected on the $n-\tau$ plane. The parameter $n$ is the number of oscillators and $\tau=g t$ is a dimensionless scaled time. The location is a straight line. }
\end{figure}
%
\section{N linearly oscillators under the influence of external force}
\noindent
In this section, we consider $n$ linearly interacting oscillators under the influence of an external field $F(t)$ applied to the first oscillator. Our aim is to investigate the propagation of the excitation araised from applying the external force along the chain. In this case the potential $-F(t)\hat{q}_1$ should be added to the Hamiltonian Eq. (\ref{chain1}). For notational simplicity we define the scaled force $R(t)=-F(t)/\sqrt{2m\hbar \omega}$. The Hamiltonian is
\begin{equation}
 \label{HF1}
  \hat{H}(t)=\sum_{j=1}^{n}\hbar \omega_j \hat{a}_{j}^\dag \hat{a}_{j}+\sum_{j=1}^{n}\hbar g\,(\hat{a}_{j} \hat{a}_{j+1}^\dag + \hat{a}_{j}^\dag \hat{a}_{j+1})\,+\,\hbar R(t)(\hat{a}_1+\hat{a}_{1}^\dag),
\end{equation}
which can be written in matrix form as
\begin{eqnarray}\label{chain29}
\hat{H}(t) &=&
\left(
\begin{array}{cccc}
	\hat{a}^{\dag}_1 & \hat{a}^{\dag}_2 &...& \hat{a}^{\dag}_n\\
\end{array}
\right)
\underbrace{
\left(
\begin{array}{cccccc}
	 \hbar\omega_0 & \hbar g & 0 & 0 & \cdots & 0  \\
	 \hbar g & \hbar\omega_0 & \hbar g & 0 & \cdots & 0 \\
      0  & \hbar g & \hbar\omega_0 & \hbar g & \dots & 0 \\
      \vdots    & \cdots     & \ddots  & \ddots & \ddots & \vdots \\
      0 & \cdots & 0 & \hbar g & \hbar\omega_0 & \hbar g \\
      0  & \cdots & 0 & 0 & \hbar g & \hbar\omega_0\\
\end{array}
\right)}_{\Gamma}
\left(
\begin{array}{c}
	\hat{a}_1 \\
    \hat{a}_2 \\
    \vdots\\
	\hat{a}_n
\end{array}
\right)\nn\\
&& \,\,\,\,\,+\hbar R(t)(\hat{a}^{\dag}_1+\hat{a}_1).
\end{eqnarray}
Let $M$ be the $n\times n$ matrix that diagonalizes the coefficient matrix $\Gamma$, $M^T\,\Gamma\,M=diag(\hbar\lambda_1,\cdots,\hbar\lambda_n)$, where $\hbar\lambda_i$ are the eigenvalues of the matrix $\Gamma$. We define the new ladder operators $\hat{B}_i$ and $\hat{B}_i^\dag$ by Bogoliubov transformations $\hat{a}_i=\sum\limits_{j=1}^n M_{ij}\,\hat{B}_j$ and $\hat{a}^\dag_i=\sum\limits_{j=1}^n M_{ij}\,\hat{B}^\dag_j$. The Hamiltonian Eq. (\ref{HF1}) in terms of the new operators $\hat{B}_i$ and $\hat{B}_i^\dag$ is written as
\begin{equation}\label{hj}.
  \hat{H}=\sum_{j=1}^n \hat{H}_j=\sum_{j=1}^n \Big[\hbar \lambda_j \hat{a}^{\dag}_j \hat{a}_j+\hbar R(t)(M_{1j}\hat{a}^{\dag}_j+M_{1j}\hat{a}_j)\Big].
\end{equation}
If $\hat{U}_j (t)$ is the evolution operator corresponding to the Hamiltonian $\hat{H}_j$, then the evolution operator corresponding to $\hat{H}$ is the tensor product
$\hat{U}(t)=\prod_{j=1}^n \bigotimes \hat{U}_j(t)$. To find $\hat{U}_j (t)$, we use the Lie algebra properties and use the ansatz \cite{shareef2022quantum}
\begin{eqnarray}\label{Ansatz}
  \hat{U}_j(t) &=& \exp(-if_0(t)/\hbar)\exp(-if_1(t) \hat{B}^{\dag}_j \hat{B}_j/\hbar)\exp(-if_2(t) \hat{B}^{\dag}_j /\hbar)\nn\\
               && \times\, \exp(-i\bar{f}_2(t) \hat{B}_j/\hbar).
\end{eqnarray}
By inserting Eq. (\ref{Ansatz}) into $i\hbar\,\partial_t\,\hat{U}_j (t)=\hat{H}_j\,\hat{U}_j (t)$, we find the unknown functions as $f_0(t)=|-\hbar M_{1j}|^2\,|\tilde{R}(t)|^2/2$, $f_1(t)=\hbar \lambda_j t$, $f_2(t)=\hbar M_{1j} \tilde{R}(t)$, and $\tilde{R}(t)=\int_{0}^{t}dt' R(t') \exp(i\lambda_j t')$. Therefore,
\begin{eqnarray}
\hat{U}_j(t) &=& e^{-\frac{1}{2}|M_{1j}|^2\,|\tilde{R}(t)|^2}\,e^{-i\lambda_j t \hat{B}^{\dag}_j \hat{B}_j}\,e^{-iM_{1j}\tilde{R}(t)\hat{B}^{\dag}_j} \,e^{-i\overline{M}_{1j}\tilde{R}(t)\hat{B}_j},\nn\\
             &=& e^{-i\lambda_j t \hat{B}^{\dag}_j \hat{B}_j}\,\hat{D}(-i M_{1j}\,\tilde{R}(t)),
\end{eqnarray}
where $\hat{D}(-i M_{1j}\,\tilde{R}(t))$ is a displacement operator that operate on vacuum stae $|0\ra_j$ and creates a coherent state $|(-i M_{1j}\,\tilde{R}(t)\ra_j$.
\subsection{Example 4:}
\noindent Let us assume that the initial state of the chain is the vacuum state $|0\ra=|0\ra_1\otimes\cdots|0\ra_n$ for the old annihilation operators $\hat{a}_i$. Since the vacuum state is unique it is also the vacuum state for the new annihilation operators $\hat{B}_i$. The evolved state is $|\psi(t)\ra=\hat{U}(t)\,|0\ra$. We have
\begin{eqnarray}
  |\psi(t)\ra &=& \prod_{j=1}^n\otimes\,\hat{U}_j (t)\,|0\ra_j,\nn \\
   &=& \prod_{j=1}^n\otimes\, e^{-i\lambda_j t \hat{a}^{\dag}_j \hat{a}_j}\,|(-i M_{1j}\,\tilde{R}(t)\ra_j,\nn\\
   &=& \prod_{j=1}^n\otimes\,|(-ie^{-i\lambda_j t} M_{1j}\,\tilde{R}(t)\ra_j.
\end{eqnarray}
Therefore,
\begin{eqnarray}
  \la \psi(t)|\hat{a}^\dag_j\hat{a}_j|0\ra &=& \sum_{k,l}^n M_{jk} \bar{M}_{jl} \la \psi(t)|\hat{B}^\dag_k\hat{B}_l|0\ra,\nn\\
   &=& |\tilde{R}(t)|^2\,\sum_{k,l}^n (\bar{M}_{jk}\,e^{it\lambda_k}\,\bar{M}_{k1}^T)(M_{jl}\,e^{-it\lambda_l}\,M_{l1}^T),\nn \\
   &=& |\tilde{R}(t)|^2\,|(e^{-\frac{i}{\hbar}t\Gamma})_{1j}|^2, \\
\end{eqnarray}
that is the mean number of excitations for the jth oscillator is the product of the transition probability and $|\tilde{R}(t)|^2$.
\section{Conclusions}
\noindent
The main result of the present paper was the closed form expression Eq. (\ref{finalK}) for the quantum propagator corresponding to the Hamiltonian Eq. (\ref{1}). We found the propagator from a scheme based on the Heisenberg equations for position and momentum operators. In the following, we found a closed form expression Eq. (\ref{evolution}) for the quantum propagator of a chain of interacting oscillators under the influence of an external force for both periodic and Dirichlet boundary conditions. Having the propagator, we can find the density matrix $\rho(\mathbf{q},\mathbf{q}',t)$ of a system described by the Hamiltonian Eq. (\ref{1}), having the initial state $\rho(\mathbf{q},\mathbf{q}',0)$, as
\begin{equation}\label{R1}
  \rho(\mathbf{q},\mathbf{q}',t)=\int\int d\mathbf{x}d\mathbf{y}\,K(\mathbf{q},t|\mathbf{x},0)\,\rho(\mathbf{x},\mathbf{y},0)\,\bar{K}(\mathbf{q}',t|\mathbf{y},0).
\end{equation}
We investigated the state and excitation propagation along a linear chain of interacting identical oscillators under the action of external source. The evolved state $|\psi(t)\ra$ of the chain was obtained for the case where initially the first oscillator was assumed to be prepared in an arbitrary state and the other oscillators were in their ground state. We also considered the action of an external force on the first oscillator and obtained a formula for excitation propagation along the chain by introducing a probability transition. The probability amplitude $P(n,\tau)$ or equivalently the probability transition between the first and the last oscillator $|\la 1|e^{-\frac{i}{\hbar}t\,\Lambda}|n\ra|^2$ was depicted in Fig. 5. From Fig. 5 it was seen that the location of the first maxima of the $P(n,\tau)$ was a straight line and to more elaborate on this, the location of the first maxima of the probability amplitude projected on $(n-\tau)$-plane, was depicted in Fig. 6, which was a straight line indicating that the speed of the first excitation propagation is a constant independent on $n$, as physically expected.

%
\section*{References}
\bibliographystyle{iopart-num}
\bibliography{references}
\end{document}